\documentclass[aps,prx,preprint,groupedaddress]{revtex4-1}
\usepackage{amsmath}
\usepackage{graphicx}
\begin{document}

%\preprint{}

\title{A full characterization of the polarization of vector light beams}


\author{Chun-Fang Li}
\email[]{cfli@shu.edu.cn}
%\thanks{}
\affiliation{Department of Physics, Shanghai University, 99 Shangda Road, 200444 Shanghai, China}


\date{\today}

\begin{abstract}

We present an approach to fully characterize the polarization of general vector light beams.
When attempting to generalize the notion of Stokes parameters to nonparaxial light beams in momentum space, we find that the Jones function that determines the Stokes parameters through the Pauli matrices is defined over a natural coordinate system that is fixed by a constant unit vector, called the Stratton vector.
We further show that the Pauli matrices represent the intrinsic degree of freedom of the polarization with respect to the natural coordinate system so that the Stratton vector acts as an additional degree of freedom that complements the intrinsic degree of freedom to fully characterize the polarization.
As a consequence of the new degree of freedom, the Stratton vector, in helicity states, a phase factor that has observable physical effects is identified.
Examples of its application to characterizing the state of polarization are also given.

\end{abstract}

%\keywords{}

\maketitle


\newpage
\section{Introduction}

It would be hard to overstate the significance of the polarization of light. The notion of light polarization, as Bj\"{o}rk et. al. \cite{Sode-BKSL} noted, comes essentially from the observation of the vibration of the electric vector of a plane light wave in the transverse plane \cite{Stra, Born-W}. The Stokes parameters \cite{Stok} that are physically observable \cite{Gold, Dama} are therefore ``the most convenient mathematical characterization for the state of polarization of a plane wave'' \cite{Jauc-R}.
In practice, however, they are commonly used to characterize the state of polarization of paraxial light beams, including the so-called scalar beams \cite{Berr-GL, Cald, Boye-LP} that are assumed to be uniformly polarized as well as the vector beams \cite{Flam-SSB, Beck-BA, Card-KSM, Turp-LPL} that vary in polarization over the transverse profile.
Such a characterization cannot be extended to the polarization of a nonparaxial light beam the electric field of which has a non-negligible component in the axial direction.
An example is the monochromatic light beam that Barnett and Allen introduced in Ref. \cite{Barn-A}, the electric field of which takes the form
\begin{equation*}
\mathbf{E}
=e^{i l \phi} \int_0^k d \kappa E(\kappa) e^{i k_z z}
\Big\{ J_l (\kappa r) (\alpha \bar{x}+\beta \bar{y})
+\frac{i \kappa}{2 k_z}
[(\alpha+i \beta)e^{-i \phi} J_{l-1} (\kappa r)
-(\alpha-i \beta)e^{i  \phi} J_{l+1} (\kappa r)] \bar{z} \Big\},
\end{equation*}
where $k_z=(k^2-\kappa^2)^{1/2}$, $J_l (\kappa r)$ is the Bessel function of order $l$, complex constants $\alpha$ and $\beta$ satisfy
$|\alpha|^2+|\beta|^2=1$, $\bar x$, $\bar y$, and $\bar z$ are the unit vectors along the corresponding coordinate axes.
As was shown before \cite{Li-WY}, the real parameter
$-i (\alpha^* \beta-\beta^* \alpha)$
is no longer the polarization ellipticity of the beam though the relative strengths of the transverse $x$- and $y$-components of the electric field are indicated by $\alpha$ and $\beta$, respectively.
Of course, one can always generalize the Stokes parameters to nonparaxial beams in momentum space on the basis of Fourier transformation. Unfortunately, it is found that so generalized Stokes parameters are not able to fully characterize the polarization of nonparaxial beams.
But thankfully, in so doing we do find out an approach to meet the need. The purpose of the present paper is to report this new approach.

Without loss of generality, we consider the electric field of an arbitrary vector beam in free space, which usually has a non-vanishing axial component. It can be written in terms of the plane-wave modes in the following way \cite{Akhi-B},
\begin{equation}\label{FT}
  \mathbf{E} (\mathbf{x},t)
 =\frac{1}{(2 \pi)^{3/2}}
  \int \mathbf{e}(\mathbf{k}) \exp[i (\mathbf{k} \cdot \mathbf{x}-\omega t)]
  d^3 k,
\end{equation}
where $\omega=c k$, $c$ is the speed of light in free space, $k=|\mathbf{k}|$, and
$\mathbf{e} (\mathbf{k})$ is the electric field in momentum space, which stands for the vector amplitude of the plane-wave mode.
Since a plane-wave mode is nothing but an eigenstate of the momentum, the notion of Stokes parameters can be strictly generalized to the general vector beam (\ref{FT}) in momentum space.
As is well known, the electric vector $\mathbf e$ at any point $\mathbf k$ in momentum space is perpendicular to the relevant momentum,
\begin{equation}\label{TC-MS}
    \mathbf{k} \cdot \mathbf{e}(\mathbf{k})=0,
\end{equation}
by virtue of the Maxwell equation
$ \nabla \cdot \mathbf{E}(\mathbf{x},t)=0 $.
To separate out the polarization from the intensity \cite{Gold, Dama, Jauc-R, Fano}, we split the strength factor off from the momentum-space electric field by writing
\begin{equation}\label{vec-e}
    \mathbf{e}(\mathbf{k})=e(\mathbf{k}) \mathbf{a}(\mathbf{k}),
\end{equation}
where the strength factor $e(\mathbf{k})$ satisfies
$|e(\mathbf{k})|=|\mathbf{e}(\mathbf{k})|$.
So separated unit-vector function $\mathbf{a} (\mathbf{k})$, which is known in the literature \cite{Akhi-B} as the polarization vector, describes the state of polarization of the general vector beam (\ref{FT}) in momentum space. What is noteworthy is that the polarization vector is not independent of the momentum. It has to satisfy the transversality condition,
\begin{equation}\label{TC}
\mathbf{k} \cdot \mathbf{a} (\mathbf{k})=0,
\end{equation}
in accordance with Eq. (\ref{TC-MS}). By this it is meant that the notion of polarization represented by the polarization vector is not an independent degree of freedom at all though it is commonly assumed \cite{Kwia-MWZ, Mich-WZ, Gadw-GD, Pan, Fick-LRZ, Bayr-SCB, Bial-B, Nech-ELB} to have that property, implicitly or explicitly.
Even so, Eq. (\ref{TC}) allows to define the Stokes parameters for the polarization state $\mathbf{a} (\mathbf{k})$ at each momentum. The problem, as we will see, is that so defined Stokes parameters need to be specified with respect to some natural coordinate system (NCS) in association with the momentum.
To completely determine the Stokes parameters, it is required to figure out a way to determine the transverse axes of the NCS.
We will show that this can be done by using the real constant unit vector that was introduced \cite{Stra, Gree-W, Patt-A, Davi-P, Li08} to represent various kinds of non-paraxial beams.
Unexpectedly, such a unit vector turns out to be an independent degree of freedom that is needed to fully characterize the polarization of general vector beams.

The contents of this paper are arranged as follows:
Section \ref{SP-int} generalizes the notion of the Stokes parameters to a general vector beam in momentum space. To completely determine the Stokes parameters, a real constant unit vector called Stratton vector (SV) is used to determine the transverse axes of the NCS.
It is shown in Section \ref{LRP} that any particular SV will fix a natural representation for the polarization. The polarization wavefunction in the natural representation, the Jones function, is defined over the associated NCS in the sense that the Stokes parameters it determines through the Pauli matrices \cite{Fano} are specified with respect to the same NCS.
It is further shown in Section \ref{DoFs} that in any natural representation, the Pauli matrices represent an independent degree of freedom of the polarization with respect to the associated NCS, referred to as the intrinsic degree of freedom. The SV to determine the NCS therefore serves as a second independent degree of freedom that complements the intrinsic degree of freedom to fully characterize the polarization.
A direct consequence of the new degree of freedom, the SV, in helicity states is discussed in Section \ref{Consequence}.
Section \ref{Applications} concerns the applications of the SV to the characterization of the polarization state of light beams when the value of the intrinsic degree of freedom is given. The last section concludes the paper with remarks.

\section{\label{SP-int} Stokes parameters in momentum space}

\subsection{\label{SP-PW} Brief review of the Stokes parameters for a single plane wave}

To generalize the Stokes parameters to a general vector beam in momentum space, it is instructive to briefly review what has been implicitly assumed in the definition of the Stokes parameters for a single plane wave.
Consider a plane wave of wave number $k_0$ in free space. Taking its propagation direction as the $z$ axis of the global coordinate system (GCS), its complex electric vector can be written as follows,
\begin{equation*}
\mathbf{E}_p =e_p \mathbf{a}_p \exp[i(k_0 z-\omega_0 t)],
\end{equation*}
where $\omega_0 =c k_0$, the constant $e_p$ is the amplitude, and the unit vector $\mathbf{a}_p$ is the polarization vector.
Since $\mathbf{a}_p$ is perpendicular to the $z$ axis, choosing the unit vectors, $\bar{x}$ and $\bar{y}$, along the corresponding transverse axes as the polarization bases, one has
\begin{equation*}
\mathbf{a}_p=\alpha_x \bar{x} +\alpha_y \bar{y}.
\end{equation*}
The expansion coefficients $\alpha_x$ and $\alpha_y$ make up the so-called Jones vector \cite{Jones}
$
\tilde{\alpha}_p =\bigg(\begin{array}{c}
                           \alpha_x \\
                           \alpha_y
                        \end{array}
                  \bigg)
$,
which satisfies
$\tilde{\alpha}_p^\dag \tilde{\alpha}_p =1$, where the superscript $\dag$ stands for the transposed conjugation.
The Stokes parameters are defined \cite{Dama, Fano} through the Pauli matrices
\begin{equation}\label{PM}
    \hat{\sigma}_1=\bigg(\begin{array}{cc}
                           1 &  0 \\
                           0 & -1
                         \end{array}
                   \bigg),                 \quad
    \hat{\sigma}_2=\bigg(\begin{array}{cc}
                           0 & 1 \\
                           1 & 0
                         \end{array}
                   \bigg),                 \quad
    \hat{\sigma}_3=\bigg(\begin{array}{cc}
                           0 & -i \\
                           i &  0
                         \end{array}
                   \bigg)
\end{equation}
and are determined by the Jones vector as follows,
\begin{equation*}
    s_{p,i}=\tilde{\alpha}_p^\dag \hat{\sigma}_i \tilde{\alpha}_p .
\end{equation*}

It is seen that in defining the Stokes parameters for a single plane wave, the propagation direction has been assumed to be one of the Cartesian axes of the GCS, the $z$ axis. Based on such an assumption, the unit vectors $\bar x$ and $\bar y$ along the transverse Cartesian axes are further taken as the polarization bases.
But because not all the plane-wave components of a general vector beam propagate along the $z$ axis, it is impossible to take $\bar x$ and $\bar y$ as the polarization bases for all of them.
How to determine the polarization bases for different plane-wave components will be our main concern in generalizing the Stokes parameters to a general vector beam.

\subsection{Generalization of Stokes parameters to a general vector beam}

Now let us turn our attention to the Stokes parameters of the general vector beam (\ref{FT}) in momentum space.
According to Eq. (\ref{TC}), to every point $\mathbf k$ in momentum space there corresponds a pair of mutually-perpendicular unit vectors, denoted by $\mathbf u$ and $\mathbf v$, which form, with $\mathbf k$, a right-handed Cartesian system obeying the following relations,
\begin{equation}\label{triad}
    \mathbf{u} \cdot  \mathbf{v}=0,                   \quad
    \mathbf{u} \times \mathbf{v}= \frac{\mathbf k}{k}.
\end{equation}
One can choose the unit vectors $\mathbf u$ and $\mathbf v$ as the polarization bases to expand the polarization vector $\mathbf{a}$ at the relevant momentum,
\begin{equation*}
    \mathbf{a}(\mathbf{k})=\alpha_1 \mathbf{u} +\alpha_2 \mathbf{v}.
\end{equation*}
The two-element Jones vector,
$
\tilde{\alpha}=\bigg(\begin{array}{c}
                      \alpha_1 \\
                      \alpha_2
                    \end{array}
               \bigg),
$
which consists of the expansion coefficients and obeys $\tilde{\alpha}^\dag \tilde{\alpha}=1$,
is in general a function of the momentum. We will call it the Jones function.
Introducing the matrix
\begin{equation}\label{varpi}
   \varpi \equiv (
                  \begin{array}{cc}
                     \mathbf{u} & \mathbf{v} \\
                  \end{array}
                 ),
\end{equation}
which contains the polarization bases as the column vectors, one can write the polarization vector in terms of the Jones function simply as
\begin{equation}\label{QUT1}
\mathbf{a}(\mathbf{k})=\varpi \tilde{\alpha}.
\end{equation}
The matrix $\varpi$ has the property
\begin{equation}\label{QU1}
    \varpi^\dag \varpi=I_2
\end{equation}
by virtue of Eqs. (\ref{triad}), where $I_2$ is the 2-by-2 unit matrix.
Multiplying Eq. (\ref{QUT1}) by $\varpi^\dag$ on the left and making use of Eq. (\ref{QU1}), one is able to express the Jones function in terms of the polarization vector as
\begin{equation}\label{QUT2}
    \tilde{\alpha} =\varpi^\dag \mathbf{a}.
\end{equation}

With the Jones function (\ref{QUT2}), it is straightforward to follow the procedure that is used for a single plane wave to define the Stokes parameters for the polarization state $\mathbf{a} (\mathbf{k})$ at each momentum,
\begin{equation}\label{SP}
    s_i=\tilde{\alpha}^\dag \hat{\sigma}_i \tilde{\alpha},
\end{equation}
where $\hat{\sigma}_i$ are the Pauli matrices (\ref{PM}).
It seems that so defined Stokes parameters could be used to characterize the polarization of general vector beams. Unfortunately, the Stokes parameters so far have not yet been completely determined.

\subsection{\label{ISV} Introduction of SV}

As mentioned above, for a single plane wave, one can always take the propagation direction as the $z$ axis of the GCS. In that case, it is allowed to choose the unit vectors along the $x$ and $y$ axes of the GCS as the polarization bases.
But for the general vector beam (\ref{FT}), it is no longer possible to take the propagation direction of each plane-wave component as the same $z$ axis for different plane-wave components will generally propagate in different directions.
The unit vectors along the $x$ and $y$ axes thus cannot be taken as the polarization bases for all the plane-wave components.
As a matter of fact, the transverse axes $\mathbf u$ and $\mathbf v$ at different momenta cannot always be the same as can be seen from Eqs. (\ref{triad}).
Importantly, they are arbitrary to the extent that a rotation about the relevant momentum $\mathbf k$ can be performed.
In order to determine the Stokes parameters completely, it is essential to figure out how to determine the transverse axes of the momentum-associated NCS $\mathbf{uvw}$.

Fortunately, it was first shown by Stratton \cite{Stra} and later by others \cite{Gree-W, Patt-A, Davi-P, Li08} that this may be done consistently by introducing a real constant unit vector $\mathbf I$, referred to as the SV, in the following way,
\begin{equation}\label{TA}%transverse axes
    \mathbf{u} =\mathbf{v} \times \frac{\mathbf k}{k},                \quad
    \mathbf{v} =\frac{\mathbf{I} \times \mathbf{k}}{|\mathbf{I} \times \mathbf{k}|}.
\end{equation}
As a matter of fact, it is easy to prove that any real constant unit vector $\mathbf I$ can determine, through these two equations, the transverse axes $\mathbf u$ and $\mathbf v$ that satisfy Eqs. (\ref{triad}).
By this it is meant that the Jones function given by Eqs. (\ref{QUT2}) and (\ref{varpi}) is always associated with some SV. Different SV's will determine different transverse axes, resulting in different Jones functions.
Consequently, defined by the Pauli matrices (\ref{PM}) no matter which SV the Jones function is associated with, the Stokes parameters are associated with the same SV as the Jones function is.
This in turn indicates that the Stokes parameters themselves are not able to solely characterize the polarization of general vector beams.

Since the unit vectors $\mathbf u$ and $\mathbf v$ as the polarization bases already satisfy the transversality condition (\ref{TC}), the Jones function is not constrained by such conditions. It can in principle be independent of the momentum.
On the basis of the quantum-mechanical theory \cite{Mess, Merz} about the spin of electrons, the momentum-independent Jones function should describe an independent degree of freedom of the polarization that is represented by the Pauli matrices (\ref{PM}).
To interpret the physical meaning of that degree of freedom, it is beneficial to analyze the properties of the Jones function in association with the SV.

\section{\label{LRP} Polarization wavefunction in natural representation}

\subsection{Transformation of Jones function under the change of SV}

To this end, let us first examine how the Jones function of a given polarization state depends on the choice of the SV.
Consider a different SV, say $\mathbf{I}'$. In this case, the transverse axes at momentum $\mathbf k$ take the form
\[
    \mathbf{u}' =\mathbf{v}' \times \frac{\mathbf k}{k}, \hspace{5pt}
    \mathbf{v}' =\frac{\mathbf{I}' \times \mathbf{k}}{|\mathbf{I}' \times \mathbf{k}|}.
\]
According to Eq. (\ref{QUT2}), the Jones function of the same polarization state $\mathbf a$ in association with the primed SV reads
\begin{equation}\label{JV'}
    \tilde{\alpha}'=\varpi'^{\dag} \mathbf{a},
\end{equation}
where
$\varpi'=(\begin{array}{cc}
            \mathbf{u}' & \mathbf{v}'
          \end{array}
         ).
$
As mentioned in Part \ref{ISV}, the primed transverse axes
$\mathbf{u}'$ and $\mathbf{v}'$
are related to the unprimed ones $\mathbf{u}$ and $\mathbf{v}$ by a local rotation about the relevant momentum $\mathbf k$.
Denoted by $\Phi$, the rotation angle obeys
\begin{subequations}\label{R-TA}
\begin{align}
  \mathbf{u}' & = \mathbf{u} \cos \Phi +\mathbf{v} \sin \Phi,   \\
  \mathbf{v}' & =-\mathbf{u} \sin \Phi +\mathbf{v} \cos \Phi.
\end{align}
\end{subequations}
These two equations can be combined into one single equation of the following form,
\begin{equation}\label{R-LTA1}
    \varpi'=\exp [-i(\hat{\mathbf \Sigma} \cdot \mathbf{w}) \Phi] \varpi,
\end{equation}
where
\begin{equation*}
    \hat{\Sigma}_x=\Bigg(\begin{array}{ccc}
                           0 & 0 &  0 \\
                           0 & 0 & -i \\
                           0 & i &  0
                         \end{array}
                   \Bigg),              \quad
    \hat{\Sigma}_y=\Bigg(\begin{array}{ccc}
                            0 & 0 & i \\
                            0 & 0 & 0 \\
                           -i & 0 & 0
                         \end{array}
                   \Bigg),              \quad
    \hat{\Sigma}_z=\Bigg(\begin{array}{ccc}
                           0 & -i & 0 \\
                           i &  0 & 0 \\
                           0 &  0 & 0
                         \end{array}
                   \Bigg)
\end{equation*}
are the generators of SO(3) rotation in the GCS and
$\mathbf{w}=\frac{\mathbf k}{k}$
is the unit wavevector.
Substituting Eqs. (\ref{R-LTA1}) and (\ref{QUT1}) into Eq. (\ref{JV'}), one has
\begin{equation}\label{T-JV}
    \tilde{\alpha}' =\exp \left(i \hat{\sigma}_3 \Phi \right) \tilde{\alpha},
\end{equation}
where the relation
\begin{equation}\label{HO}
    \hat{\sigma}_3
   =\varpi^\dag (\hat{\mathbf \Sigma} \cdot \mathbf{w}) \varpi ,
\end{equation}
which holds irrespective of the SV, has been used.
According to Eqs. (\ref{TA}), the rotation angle $\Phi$ in Eq. (\ref{R-LTA1}) depends not only on the SV's $ \mathbf{I}' $ and $ \mathbf{I} $ but also on the unit wavevector $\mathbf{w}$ unless
$ \mathbf{I}' =-\mathbf{I} $. In that case, one has $ \Phi=\pi $.
Eq. (\ref{T-JV}) is the transformation of the Jones function under the change of the SV.

It is noted that Eq. (\ref{HO}) reflects the correspondence between the SO(3) and SU(2) rotations \cite{Tung}.
In fact, a comparison of Eq. (\ref{T-JV}) with Eq. (\ref{JV'}) leads to
$$ \varpi'^\dag \mathbf{a}= \exp(i \hat{\sigma}_3 \Phi) \varpi^\dag \mathbf{a} $$
when Eq. (\ref{QUT2}) is taken into account. Considering the arbitrariness of the polarization vector $\mathbf a$, one must have
$\varpi'^\dag =\exp(i \hat{\sigma}_3 \Phi) \varpi^\dag$
or, equivalently,
\begin{equation}\label{R-LTA2}
    \varpi' =\varpi \exp(-i \hat{\sigma}_3 \Phi),
\end{equation}
which shows that the SO(3) rotation of the transverse axes about the momentum corresponding to the change of the SV can also be expressed by a SU(2) rotation.
In particular, the generator of the SU(2) rotation about the momentum is the third Pauli matrix $\hat{\sigma}_3$ in consistency with the correspondence (\ref{HO}) between
$\hat{\mathbf \Sigma} \cdot \mathbf{w}$ and $\hat{\sigma}_3$.
This shows that the transformation (\ref{T-JV}) is in fact a SU(2) rotation of the Jones function about the momentum.
Such a transformation implies that the Jones function in association with any particular SV is defined over the NCS that is determined by the same SV.

\subsection{Jones function is defined over NCS}

According to Eq. (\ref{SP}), the Stokes parameters in association with the primed SV are given by
\begin{equation}\label{SP'}
    s'_i =\tilde{\alpha}'^\dag \hat{\sigma}_i \tilde{\alpha}'.
\end{equation}
Upon substituting Eq. (\ref{T-JV}) and using Eq. (\ref{SP}), one gets
\begin{subequations}\label{T-SP}
\begin{align}
  s'_1 & =  s_1 \cos 2\Phi +s_2 \sin 2\Phi,  \label{SP1} \\
  s'_2 & = -s_1 \sin 2\Phi +s_2 \cos 2\Phi,  \label{SP2} \\
  s'_3 & =  s_3.                             \label{SP3}
\end{align}
\end{subequations}
These are the transformations of the Stokes parameters under the change of the SV. Only the first two Stokes parameters depend on the choice of the SV. The third one does not. The reason lies with the correspondence (\ref{HO}).
It is noted that when $ \mathbf{I}' =-\mathbf{I} $, that is to say, when $ \Phi=\pi $, one has $ s'_i =s_i $.

Traditionally, the Stokes vector \cite{Gold, Dama, Jauc-R, Fano} formed by the Stokes parameters in the case of a single plane wave is depicted on the surface of the Poincar\'{e} sphere.
But because the transformations (\ref{T-SP}) come from a rotation of the transverse axes about the momentum, the third Stokes parameter that is invariant under such a rotation should be considered as the component of the Stokes vector along the momentum. Accordingly, the first two Stokes parameters in association with any particular SV should be considered as the components of the Stokes vector along the corresponding transverse axes.
Specifically, the Stokes parameters (\ref{SP}) in association with the unprimed SV $\mathbf I$ form the following Stokes vector,
\begin{equation}\label{SV}
    \mathbf{s}=s_1 \mathbf{u} +s_2 \mathbf{v} +s_3 \mathbf{w},
\end{equation}
where
$\mathbf u$ and $\mathbf v$ are the unit vectors given by Eqs. (\ref{TA}).
Letting
\begin{equation}\label{PMV}
\hat{\boldsymbol \sigma}
=\hat{\sigma}_1 \mathbf{u} +\hat{\sigma}_2 \mathbf{v}
+\hat{\sigma}_3 \mathbf{w},
\end{equation}
one can rewrite Eq. (\ref{SV}) as
\begin{equation*}
\mathbf{s}=\tilde{\alpha}^\dag \hat{\boldsymbol \sigma} \tilde{\alpha}.
\end{equation*}
Mathematically, it states that the Jones function (\ref{QUT2}) in association with any particular SV via Eqs. (\ref{varpi}) and (\ref{TA}) is defined over the NCS that is determined by the same SV.
This is to be compared with the polarization vector that is defined over the GCS.
In a word, the Stokes vector of a polarization state is always associated with some particular SV.
What is noteworthy is that the Stokes vectors of the same polarization state in association with different SV's are in general not the same. To see this, let us look at the Stokes vector in association with the primed SV,
\begin{equation*}
    \mathbf{s}'=\tilde{\alpha}'^\dag \hat{\boldsymbol \sigma}' \tilde{\alpha}'
               =s'_1 \mathbf{u}' +s'_2 \mathbf{v}' +s'_3 \mathbf{w},
\end{equation*}
where
$\hat{\boldsymbol \sigma}' =\hat{\sigma}_1 \mathbf{u}' +\hat{\sigma}_2 \mathbf{v}' +\hat{\sigma}_3 \mathbf{w}$.
Upon substituting Eqs. (\ref{T-SP}) and (\ref{R-TA}), one finds
\begin{equation}\label{T-SV}
    \mathbf{s}'
   =\exp [i(\hat{\mathbf \Sigma} \cdot \mathbf{w}) \Phi] \mathbf{s}.
\end{equation}
This can be explained in terms of their transverse components
$\mathbf{s}_\perp =s_1 \mathbf{u} +s_2 \mathbf{v}$
and
$\mathbf{s}'_\perp =s'_1 \mathbf{u}' +s'_2 \mathbf{v}'$
as follows.

It is well known that the Pauli matrices (\ref{PM}) satisfy the SU(2) algebra,
\begin{equation}\label{CCR}
    \Big[ \frac{\hat{\sigma}_i}{2}, \frac{\hat{\sigma}_j}{2} \Big]
   =i \sum_k \varepsilon_{ijk} \frac{\hat{\sigma}_k}{2},
\end{equation}
where $\varepsilon_{ijk}$ is the Levi-Civit\'{a} pseudotensor. In view of this, it follows from Eq. (\ref{PMV}) that each of them, except for a factor $\frac{1}{2}$, is the generator of a SU(2) rotation about the corresponding Cartesian axes of the NCS. This again shows that the third Pauli matrix $\hat{\sigma}_3$ is the generator of a SU(2) rotation about the momentum.
It is thus seen from Eq. (\ref{T-JV}) that the primed Jones function is the result of the rotation of the unprimed Jones function about the momentum by an angle of $-2 \Phi$.
If the transverse axes of the NCS were not rotated, the first two primed Stokes parameters $s'_1$ and $s'_2$ would form a transverse component,
$\mathbf{s}''_\perp =s'_1 \mathbf{u} +s'_2 \mathbf{v}$,
which is equal to the result of the rotation of $\mathbf{s}_\perp$ about the momentum by the same angle $-2 \Phi$. This is what Eqs. (\ref{SP1}) and (\ref{SP2}) show and is graphically displayed in Fig. 1(a).
\begin{figure}[tb]
	\centerline{\includegraphics[width=11cm]{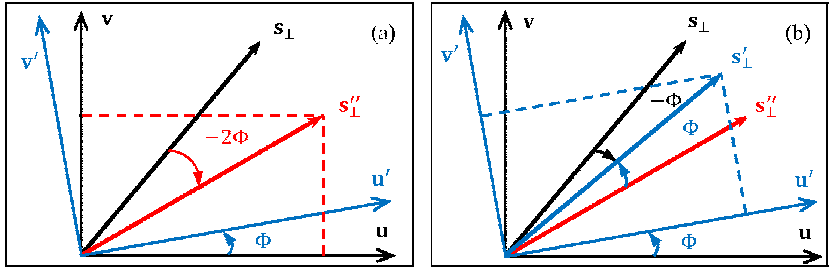}}
	\caption{(a) $\mathbf{s}''_\perp$ is the result of the rotation of $\mathbf{s}_\perp$ by an angle of $-2 \Phi$. (b) $\mathbf{s}''_\perp$ is rotated to $\mathbf{s}'_\perp$ along with the unprimed NCS being rotated to the primed NCS by an angle of $\Phi$.}
\end{figure}
However, as mentioned above, the primed transverse axes $\mathbf{u}'$ and $\mathbf{v}'$ result from the rotation of the unprimed transverse axes $\mathbf{u}$ and $\mathbf{v}$ about the momentum by the angle $\Phi$.
Along with the unprimed NCS being rotated to the primed NCS, the transverse component $\mathbf{s}''_\perp$ is rotated to $\mathbf{s}'_\perp$ as is displayed in Fig. 1(b). As a consequence, $\mathbf{s}'_\perp$ is equal to the result of the rotation of $\mathbf{s}_\perp$ about the momentum by an angle of $-\Phi$. This is just what Eq. (\ref{T-SV}) means.
By the way, it is pointed out that the primed Stokes vector is different from the unprimed one even when
$ \mathbf{I}' =-\mathbf{I} $, in contrast with the transformations (\ref{T-SP}) on the Stokes parameters.

In short, the Jones function (\ref{QUT2}) in association with one particular SV is defined over the NCS that is fixed exactly by that SV. The Stokes parameters it determines are therefore specified with respect to the same NCS, forming the Stokes vector (\ref{SV}).
It is worth noting that what we conclude here is valid for any kind of light beams. As a corollary, the Stokes parameters reviewed in Part \ref{SP-PW} for a single plane wave are specified with respect to the GCS $xyz$ as Yun et. al. \cite{Yun-CC} realized recently.

\subsection{SV fixes a natural representation for polarization}

It is now clear that any SV will determine, through Eq. (\ref{varpi}), a matrix $\varpi$ that obeys Eq. (\ref{QU1}). With the help of this property, one readily gets from Eq. (\ref{QUT1})
\begin{equation*}
    \mathbf{a}^\dag \mathbf{a}=\alpha^\dag \alpha.
\end{equation*}
Upon substituting Eq. (\ref{QUT2}) and considering the arbitrariness of the polarization vector $\mathbf a$ into account, one immediately arrives at
\begin{equation}\label{QU2}
    \varpi \varpi^\dag=I_3 ,
\end{equation}
where $I_3$ is the 3-by-3 unit matrix.
What is conveyed by Eqs. (\ref{QU1}) and (\ref{QU2}) is the fact that the matrix $\varpi$ in association with any SV is a quasi-unitary matrix \cite{Golu-L}. $\varpi^\dag$ is the Moore-Penrose pseudo inverse of $\varpi$, and vice versa.
This means that any SV will fix a quasi-unitary transformation between the polarization vector over the GCS and the Jones function over the associated NCS via Eqs. (\ref{varpi}) and (\ref{QUT2}) or its inverse transformation via Eqs. (\ref{varpi}) and (\ref{QUT1}).
Specifically, to each polarization vector $\mathbf a$ over the GCS there corresponds one unique Jones function over the associated NCS via Eq. (\ref{QUT2}). Conversely, to each Jones function $\tilde \alpha$ over the NCS that is determined by one particular SV, there corresponds one unique polarization vector via Eq. (\ref{QUT1}).
In the language of quantum mechanics, any SV will fix, via Eqs. (\ref{varpi}) and (\ref{QUT2}), a representation for the polarization.
Remarkably, this is a representation that is different from the global representation in which the state of polarization is described by the polarization vector over the GCS.
It is a natural representation. The wavefunction of the polarization, the Jones function, in it is defined over the associated NCS;
the physically observable quantities, the Stokes parameters, that are represented by the Pauli matrices are specified with respect to the same NCS as is expressed by Eq. (\ref{SV}).

More importantly, due to the fact that the quasi unitarity of the transformation matrix $\varpi$ holds irrespective of the SV, there are an infinite number of natural representations for the polarization. Each SV will fix a different natural representation.
All the natural representations are equivalent in the sense that a polarization state can be described in any natural representation by a corresponding wavefunction. The wavefunctions in different natural representations are not the same. They are related to one another by Eq. (\ref{T-JV}).
But on the other hand, all the natural representations are not equivalent in the sense that a Jones function in different natural representations does not mean the same state of polarization as can be seen from Eq. (\ref{QUT1}).
This indicates that the SV is an independent degree of freedom of the polarization.

\section{\label{DoFs} Two independent degrees of freedom of polarization}

\subsection{Intrinsic degree of freedom with respect to NCS}

The wavefunction (\ref{QUT2}) of a polarization state $\mathbf a$ in one particular natural representation is usually dependent on the momentum. So are the Stokes parameters (\ref{SP}) it determines through the Pauli matrices (\ref{PM}).
Nevertheless, as emphasized before, the polarization wavefunction in any natural representation is not constrained by such conditions as Eq. (\ref{TC}).
According to the quantum-mechanical theory \cite{Mess, Merz} about the spin of electrons, the Pauli matrices in any natural representation represent an independent degree of freedom of the polarization,
referred to as the intrinsic degree of freedom.
The intrinsic characteristic here is two-fold \cite{Li-Z}. Firstly, the Pauli matrices (\ref{PM}) as constant operators represent physical observables that are independent of such variables as the momentum. Secondly, as is explicitly expressed by Eq. (\ref{PMV}), these observables are specified with respect to the NCS rather than with respect to the GCS.

To demonstrate this in more detail, we consider an arbitrary constant Jones function $\tilde \nu $, which is the eigenfunction of the constant matrix
$ n_1 \hat{\sigma}_1 +n_2 \hat{\sigma}_2 +n_3 \hat{\sigma}_3 \equiv n_i \hat{\sigma}_i $,
\begin{equation*}
(n_i \hat{\sigma}_i) \tilde{\nu} =\tilde{\nu},
\end{equation*}
where the constants $ n_i = \tilde{\nu}^\dag \hat{\sigma}_i \tilde{\nu} $ are the Stokes parameters.
On one hand, the physical meaning of the constant Jones function $\tilde \nu$ is not clear if the natural representation in which it is defined is indeterminate.
In fact, according to Eq. (\ref{QUT1}), the state of polarization it describes in natural representation $\mathbf I$ is given by
\begin{equation*}
\mathbf{n} = \varpi \tilde{\nu}.
\end{equation*}
The constant Stokes parameters $n_i$ are the momentum-independent observables of this polarization state with respect to the associated NCS $\mathbf{uvw}$, forming the Stokes vector
$ n_1 \mathbf{u} +n_2 \mathbf{v} +n_3 \mathbf{w} $.
But on the other hand, remember that different natural representations are equivalent. Denoting by $\tilde{\nu}'$ the Jones function of the same polarization state $\mathbf n$ in a different natural representation, say $\mathbf{I}'$, one has
\begin{equation*}
    \tilde{\nu}' =\exp(i \hat{\sigma}_3 \Phi) \tilde{\nu}
\end{equation*}
in accordance with Eq. (\ref{T-JV}). Unexpectedly, it is in general not the eigenfunction of the matrix
$ n_i \hat{\sigma}_i $. In particular, it becomes dependent on the momentum. So do the Stokes parameters with respect to the primed NCS as can be seen from Eqs. (\ref{T-SP}).
This means that only in some particular natural representation can the wavefunction of a polarization state be constant. After all, what the Pauli matrices represent in a natural representation is only the degree of freedom of the polarization with respect to the associated NCS.
The term \textit{intrinsic} is more appropriate for such a kind of degree of freedom.

\subsection{SV is another degree of freedom of polarization}

Now that the Pauli matrices (\ref{PM}) only represent the intrinsic degree of freedom of the polarization with respect to the NCS, the SV that determines the NCS is another independent degree of freedom of the polarization.
That is to say, there are two independent mechanisms to change the polarization state of a light beam.
One is to change the value of the intrinsic degree of freedom, or the Jones function, with the SV remaining fixed as can be seen from Eq. (\ref{QUT1}). This mechanism can be expressed mathematically by a SU(2) rotation of the Jones function that is generated by the Pauli matrices within a fixed NCS \cite{Jauc-R}.
As an example, we consider the normalized eigenfunction of the first Pauli matrix $\hat{\sigma}_1$,
$
\tilde{\alpha}_{1+} = \bigg(\begin{array}{c}
                               1 \\
                               0
                            \end{array}
                      \bigg)
$,
belonging to the eigenvalue $+1$.
According to Eq. (\ref{QUT1}), the polarization state it describes in natural representation $\mathbf I$ is given by
\begin{equation}\label{a1}
\mathbf{a}_{1+} =\varpi \tilde{\alpha}_{1+} =\mathbf{u}.
\end{equation}
If the Jones function is changed from $\tilde{\alpha}_{1+}$ to
$
\tilde{\alpha}_{1-} =\bigg(\begin{array}{c}
                              0 \\
                              1
                           \end{array}
                     \bigg),
$
the eigenfunction of $ \hat{\sigma}_1 $ belonging to the eigenvalue $-1$, the polarization state will become
\begin{equation*}
\mathbf{a}_{1-} =\varpi \tilde{\alpha}_{1-} =\mathbf{v}.
\end{equation*}
It is, of course, different from the polarization state $\mathbf{a}_{1+}$. In fact, it is orthogonal to $\mathbf{a}_{1+}$.
Such a change can be expressed by a SU(2) rotation of $\tilde{\alpha}_{1+}$ about the momentum by the angle of $\pi$ within the same natural representation,
\begin{equation}\label{R-JF1}
\tilde{\alpha}_{1-}
=\exp \bigg(-i \frac{\hat{\sigma}_3}{2} \pi \bigg) \tilde{\alpha}_{1+} .
\end{equation}
In other words, the polarization state $\mathbf{a}_{1-}$ can be expressed in terms of the eigenfunction $\tilde{\alpha}_{1+}$ as follows,
\begin{equation}\label{a2}
\mathbf{a}_{1-}
=\varpi \exp \bigg(-i \frac{\hat{\sigma}_3}{2} \pi \bigg) \tilde{\alpha}_{1+} .
\end{equation}
The other mechanism is to change the NCS, or the SV, with the Jones function remaining fixed. To illustrate this, we consider the above-mentioned eigenfunction $ \tilde{\alpha}_{1+} $ in a different natural representation, say $\mathbf{I}'$. In this case, the polarization state it describes is given by
\begin{equation}\label{a1-P}
    \mathbf{a}'_{1+} =\varpi' \tilde{\alpha}_{1+} =\mathbf{u}',
\end{equation}
which is different from $ \mathbf{a}_{1+} $, embodying that a Jones function in different natural representations describes different states of polarization.
Substituting Eq. (\ref{R-LTA1}) into Eq. (\ref{a1-P}) and considering Eq. (\ref{a1}), one has
\begin{equation*}
    \mathbf{a}'_{1+} =\exp[-i(\hat{\mathbf \Sigma} \cdot \mathbf{w}) \Phi] \mathbf{a}_{1+} .
\end{equation*}
This is a SO(3) rotation of the polarization vector $\mathbf{a}_{1+} $ about the momentum.

To see that the latter mechanism is independent of the former one, we make use of the correspondence (\ref{HO}) and substitute Eq. (\ref{R-LTA2}) into Eq. (\ref{a1-P}) to get
\begin{equation*}
    \mathbf{a}'_{1+}
   =\varpi \exp (-i \hat{\sigma}_3 \Phi) \tilde{\alpha}_{1+} .
\end{equation*}
A comparison with Eq. (\ref{a2}) shows that changing the SV of the polarization state $\mathbf{a}_{1+} $ is not able to get the polarization state $\mathbf{a}_{1-} $. After all, the angle of the SO(3) rotation about the momentum, which obeys Eq. (\ref{R-LTA1}), is dependent on the momentum.
In a word, there are two independent physical mechanisms to change the state of polarization of a light beam.
One is expressed by a SU(2) rotation of the Jones function in a fixed NCS. Since the Jones function is not constrained by such conditions as Eq. (\ref{TC}), the SU(2) rotation may be arbitrary.
The other is expressed by a SO(3) rotation of the polarization vector in the GCS. Due to the constraint of condition (\ref{TC}) on the polarization vector, the SO(3) rotation must be around the momentum by the angle $\Phi$ that satisfies Eq. (\ref{R-LTA1}).
Concrete examples of both mechanisms will be given in Section \ref{Applications}.

From these results we conclude that the polarization of a light beam is not purely intrinsic. It cannot be solely characterized by the intrinsic quantum number that is determined by the canonical commutation relation (\ref{CCR}). A complimentary degree of freedom is needed.
This might reveal the physics beyond the transversality condition (\ref{TC}) that is expressed in terms of the polarization vector.
It is worth pointing out that different from the intrinsic degree of freedom that is represented by Hermitian operators, the Pauli matrices, the complimentary degree of freedom is a unit vector, the SV.
The newly identified degree of freedom enables us to identify a phase factor that has observable physical effects.

\section{\label{Consequence} One consequence of the new degree of freedom}

It is now clear that when the natural representation is fixed by one particular SV, each polarization state is described by a unique wavefunction via Eq. (\ref{QUT2}). Its Stokes parameters with respect to the associated NCS form the Stokes vector (\ref{SV}).
Under the change of the natural representation, the wavefunction is transformed according to Eq. (\ref{T-JV}) and the Stokes vector is transformed according to Eq. (\ref{T-SV}).
From these results one might speculate that the notion of Stokes vector in a fixed natural representation is able to solely characterize the polarization state of general vector beams.
Regrettably, this is again not true. The reason lies with the fact that the correspondence (\ref{HO}) holds irrespective of the concrete SV.

Here we are concerned about the eigenfunction of the Pauli matrix $\hat{\sigma}_3$,
$
\tilde{\alpha}_{3 \pm} =\frac{1}{\sqrt 2} \bigg(
                                            \begin{array}{c}
                                                  1 \\
                                              \pm i \\
                                            \end{array}
                                          \bigg).
$
It satisfies the eigenvalue equation
\begin{equation}\label{EVE}
    \hat{\sigma}_3 \tilde{\alpha}_{3 \pm} =\pm \tilde{\alpha}_{3 \pm}.
\end{equation}
Let us compare the states of polarization it describes in two different natural representations. In the unprimed natural representation, it describes the following polarization state,
\begin{equation}\label{a3}
    \mathbf{a}_{3 \pm} =\varpi \tilde{\alpha}_{3 \pm} =\frac{1}{\sqrt 2}(\mathbf{u} \pm i \mathbf{v}).
\end{equation}
Similarly, the polarization state it describes in the primed natural representation is given by
\begin{equation*}
    \mathbf{a}'_{3 \pm} =\varpi' \tilde{\alpha}_{3 \pm}
                        =\frac{1}{\sqrt 2} (\mathbf{u}' \pm i \mathbf{v}').
\end{equation*}
But upon substituting Eq. (\ref{R-LTA2}) and considering Eqs. (\ref{EVE}) and (\ref{a3}) into account, one finds
\begin{equation}\label{a3-P}
    \mathbf{a}'_{3 \pm} =\mathbf{a}_{3 \pm} \exp(\mp i \Phi).
\end{equation}
It differs from $\mathbf{a}_{3 \pm}$ only by the phase factor $\exp(\mp i \Phi)$. This is explained as follows.
Remembering that the Pauli matrix $\hat{\sigma}_3$ is the helicity operator \cite{Jauc-R} in the natural representation, Eq. (\ref{HO}) implies that
$\hat{\mathbf \Sigma} \cdot \mathbf{w}$
is the helicity operator in the global representation.
In fact, because Eq. (\ref{QUT2}) acts as the transformation of the polarization vector in the global representation into the polarization wavefunction in the natural representation, Eq. (\ref{HO}) plays the role of transforming the global-representation helicity operator to the natural-representation one.
With the help of Eqs. (\ref{QU1}) and (\ref{QU2}), one readily obtains from Eq. (\ref{HO})
\begin{equation*}
    \hat{\mathbf \Sigma} \cdot \mathbf{w} =\varpi \hat{\sigma}_3 \varpi^\dag.
\end{equation*}
It transforms the natural-representation helicity operator back to the global-representation one.
Now that this transformation holds irrespective of the SV, the two different polarization states $\mathbf{a}_{3 \pm}$ and $\mathbf{a}'_{3 \pm}$ are both the eigenstates of $\hat{\mathbf \Sigma} \cdot \mathbf{w}$ belonging to the same eigenvalue $\pm 1$,
\begin{equation*}
    (\hat{\mathbf \Sigma} \cdot \mathbf{w}) \mathbf{a}_{3 \pm}  =\pm \mathbf{a}_{3 \pm} , \quad
    (\hat{\mathbf \Sigma} \cdot \mathbf{w}) \mathbf{a}'_{3 \pm} =\pm \mathbf{a}'_{3 \pm} ,
\end{equation*}
by virtue of Eqs. (\ref{QU1}) and (\ref{EVE}).

On the other hand, as mentioned before, different natural representations are equivalent. Denoting by $\tilde{\alpha}'_{3 \pm}$ the wavefunction of the helicity state $\mathbf{a}_{3 \pm}$ in the primed natural representation,
one finds by use of Eqs. (\ref{T-JV}) and (\ref{EVE})
\begin{equation}\label{RT-alpha3}
\tilde{\alpha}'_{3 \pm} =\exp(i \hat{\sigma}_3 \Phi) \tilde{\alpha}_{3 \pm}
=\tilde{\alpha}_{3 \pm} \exp(\pm i \Phi).
\end{equation}
It is only different from $\tilde{\alpha}_{3 \pm}$ by a phase factor. The Stokes vector it determines is therefore the same as $\tilde{\alpha}_{3 \pm}$ does in the primed natural representation. They are all $\pm \mathbf{w}$.
This shows that the helicity states $\mathbf{a}_{3 \pm}$ and $\mathbf{a}'_{3 \pm}$ in association with two different SV's share the same Stokes vector in the same natural representation.
In other words, the two different polarization states $\mathbf{a}_{3 \pm}$ and $\mathbf{a}'_{3 \pm}$ cannot be distinguished from each other by their Stokes vectors in the same natural representation.
This is understandable. After all, the Stokes vector is a quantity that is defined in the natural representation, whereas the Stokes vector of the helicity state does not convey any information about the natural representation.

However, as we have seen, the angle $\Phi$ of the SO(3) rotation corresponding to the change of the SV depends on the momentum. So the helicity-dependent phase factor $\exp(\mp i \Phi)$ in Eq. (\ref{a3-P}) will have an impact on the spatial profile of the light beam in accordance with Eqs. (\ref{vec-e}) and (\ref{FT}).
That is to say, the polarization vectors $\mathbf{a}_{3 \pm}$ and $\mathbf{a}'_{3 \pm}$ will produce two different electric fields in position space for the same strength factor $e(\mathbf{k})$. Such an impact usually results in a helicity-dependent transverse shift of the barycenter of the beam \cite{Li08}.
For example, the so-called spin Hall effect of light that Hosten and Kwiat \cite{Host-K} experimentally observed in the diffraction at an interface between two dielectric media was quantitatively interpreted \cite{Li09b} in terms of the deflection of the SV though it was originally interpreted in terms of the helicity-dependent geometric phase in momentum space. This can be understood as follows.
First of all, the incident beam in that experiment can be treated \cite{Host-K} as a uniformly-polarized paraxial beam, which means that it has a SV that is perpendicular to its propagation direction \cite{Li08}. Secondly, the diffraction at the interface was shown \cite{Li09b} to amount to, in the first-order paraxial approximation, deflecting the SV of the beam to a direction that is not perpendicular to the propagation direction.
So the geometric phase that Hosten and Kwiat deduced in their article is nothing but the helicity-dependent phase that reflects the deflection of the SV as is clearly expressed by Eq. (\ref{a3-P}).
As a matter of fact, the effect of the SV on the transverse shift of the beam's barycenter can be extremely large in comparison with its wavelength under reasonable conditions \cite{Li07b, Yang-L}.

\section{\label{Applications} Applications of the new degree of freedom}

At last, let us make use of two concrete examples to appreciate how the SV affects the polarization state of a light beam when the value of the intrinsic degree of freedom is given.
We will see that the new degree of freedom makes it possible to describe paraxial beams of uniform and cylindrically-symmetric polarizations in a unified framework.

To this end, we assume that the light beams under investigation propagate along the $z$ axis.
In the first example, the SV is set to be perpendicular to the propagation axis \cite{Patt-A}, $\mathbf{I}^a =-\bar{x}$.
The transverse axes of the NCS that it determines take the form
\begin{eqnarray*}
  \mathbf{u}^a &=& \frac{1}{C}
                   \bigg(\bar{x}-\frac{k_\rho^2}{k^2} \bar{\rho} \cos \varphi
                        -\frac{k_z k_\rho}{k^2} \bar{z} \cos \varphi \bigg), \\
  \mathbf{v}^a &=& \frac{1}{C} \bigg(\frac{k_z}{k} \bar{y}
                                    -\frac{k_\rho}{k} \bar{z} \sin \varphi \bigg),
\end{eqnarray*}
where $\bar \rho$ stands for the radial unit vector in cylindrical coordinates, $k_z$ and $k_\rho$ are the axial and radial components of $\mathbf k$, respectively, $\varphi$ is the azimuthal angle, and
\begin{equation*}
    C=\bigg(1-\frac{k_\rho^2}{k^2} \cos^2 \varphi \bigg)^{1/2}.
\end{equation*}
(i) The first state of polarization that we are concerned with is described by $\tilde{\alpha}_{1+} $, one of the eigenfunctions of the Pauli matrix $\hat{\sigma}_1$,
\begin{equation*}
    \mathbf{a}_{1+}^a
   =\varpi^a \tilde{\alpha}_{1+}
   =\frac{1}{C} \bigg(\bar{x}-\frac{k_\rho^2}{k^2} \bar{\rho} \cos \varphi
                     -\frac{k_z k_\rho}{k^2} \bar{z} \cos \varphi \bigg),
\end{equation*}
where
$\varpi^a =(\begin{array}{cc}
              \mathbf{u}^a & \mathbf{v}^a
            \end{array}
           )$.
(ii) The second state of polarization is described by $\tilde{\alpha}_{1-} $, the other eigenfunction of $\hat{\sigma}_1$,
\begin{equation*}
    \mathbf{a}_{1-}^a
   =\varpi^a \tilde{\alpha}_{1-}
   =\frac{1}{C} \bigg(\frac{k_z}{k} \bar{y}
                     -\frac{k_\rho}{k} \bar{z} \sin \varphi \bigg),
\end{equation*}
which is orthogonal to $\mathbf{a}_{1+}^a$. It is observed that in the zeroth-order paraxial approximation in which $k_z \approx k$ and $\frac{k_\rho}{k} \approx 0$, $\mathbf{a}_{1+}^a$ reduces to $\bar x$ and $\mathbf{a}_{1-}^a$ reduces to $\bar y$, giving rise to uniformly-distributed $x$ and $y$ polarizations in position space, respectively \cite{Li08, Wang-YL}.
But it should be emphasized that the polarization state $\mathbf{a}_{1-}^a$ cannot be obtained by rotating the polarization state $\mathbf{a}_{1+}^a$ about the $z$ axis.
In fact, as is shown by Eq. (\ref{R-JF1}), it results from the SU(2) rotation of the Jones function $\tilde{\alpha}_{1+}$ of the polarization state $\mathbf{a}_{1+}^a$ about the momentum by the angle $\pi$ with the SV $\mathbf{I}^a$ remaining fixed.
(iii) The third state of polarization is described by $\tilde{\alpha}_{3+}$, one of the eigenfunctions of $\hat{\sigma}_3$,
\begin{equation*}
    \mathbf{a}_{3+}^a
   =\varpi^a \tilde{\alpha}_{3+}
   =\frac{1}{\sqrt{2} C} \bigg[ \bar{x}+i\frac{k_z}{k} \bar{y}
   -\frac{k_\rho^2}{k^2} \bar{\rho} \cos \varphi -\frac{k_\rho}{k}
    \bigg(\frac{k_z}{k} \cos\varphi+i \sin\varphi \bigg)\bar{z} \bigg].
\end{equation*}
In the zeroth-order paraxial approximation, it reduces to
$\frac{1}{\sqrt{2}}(\bar{x}+i \bar{y})$,
which gives rise to the uniformly-distributed circular polarization in position space.
It is noted that the polarization state $\mathbf{a}_{3+}^a$ can be obtained by a SU(2) rotation of the Jones function $\tilde{\alpha}_{1+}$ of the polarization state $\mathbf{a}_{1+}^a$ about the natural coordinate axis $\mathbf{v}$ by the angle $-\frac{\pi}{2}$,
\begin{equation*}
    \tilde{\alpha}_{3+}
   =\exp \Big[-i \frac{\hat{\sigma}_2}{2} \Big( -\frac{\pi}{2} \Big) \Big] \tilde{\alpha}_{1+} ,
\end{equation*}
with the SV $\mathbf{I}^a$ remaining fixed.

In the second example, the SV is set to be directed along the propagation axis of the beam \cite{Davi-P},
$\mathbf{I}^b =\bar z$.
The transverse axes it determines are given by
\begin{eqnarray*}
  \mathbf{u}^b &=& \frac{k_z}{k} \bar{\rho}-\frac{k_\rho}{k} \bar{z}, \\
  \mathbf{v}^b &=& \bar{\varphi},
\end{eqnarray*}
where $\bar{\varphi}$ stands for the azimuthal unit vector in cylindrical coordinates.
(i) In this case, the polarization state described by the Jones function
$\tilde{\alpha}_{1+} $ becomes
\begin{equation*}
    \mathbf{a}_{1+}^b =\varpi^b \tilde{\alpha}_{1+}
                   =\frac{k_z}{k} \bar{\rho}-\frac{k_\rho}{k} \bar{z},
\end{equation*}
where
$\varpi^b =(\begin{array}{cc}
              \mathbf{u}^b & \mathbf{v}^b
            \end{array}
           )$.
In the zeroth-order paraxial approximation, it will give rise to radially polarized beams through Eqs. (\ref{vec-e}) and (\ref{FT}) when the strength factor $e(\mathbf{k})$ is rotationally invariant about the $z$ axis \cite{Li07}.
(ii) In addition, the polarization state described by the Jones function
$\tilde{\alpha}_{1-} $ is given by
\begin{equation*}
    \mathbf{a}_{1-}^b =\varpi^b \tilde{\alpha}_{1-} =\bar{\varphi},
\end{equation*}
which is orthogonal to $\mathbf{a}_{1+}^b$. It will give rise to azimuthally polarized beams through Eqs. (\ref{vec-e}) and (\ref{FT}) when the strength factor $e(\mathbf{k})$ is rotationally invariant about the $z$ axis.
(iii) Moreover, the polarization state described by the Jones function $\tilde{\alpha}_{3+}$ takes the form
\begin{equation*}
    \mathbf{a}_{3+}^b
   =\varpi^b \tilde{\alpha}_{3+}
   =\frac{1}{\sqrt 2} \Big(\frac{k_z}{k} \bar{\rho} +i \bar{\varphi}
                          -\frac{k_\rho}{k} \bar{z} \Big).
\end{equation*}
It is noted that this polarization vector reduces to
$\frac{1}{\sqrt 2} (\bar{\rho} +i \bar{\varphi})
=\frac{1}{\sqrt 2} (\bar{x}+i \bar{y}) \exp (-i \varphi)$
in the zeroth-order paraxial approximation, which is different from the zeroth-order paraxial approximation of $\mathbf{a}_{3+}^a$ by the phase factor
$\exp (-i \varphi)$.

Clearly, the polarization states $\mathbf{a}_{1+}^a $ and $\mathbf{a}_{1+}^b $ are different from each other though they share the same Jones function. They differ by their SV's $\mathbf{I}^a$ and $\mathbf{I}^b$.
This is true of the polarization states $\mathbf{a}_{1-}^a $ and $\mathbf{a}_{1-}^b $ as well as the polarization states $\mathbf{a}_{3+}^a $ and $\mathbf{a}_{3+}^b $.

\section{\label{CandR} Conclusions and remarks}

In conclusion, it is revealed that two independent degrees of freedom are needed to fully characterize the polarization of general vector beams. One is the SV that determines the transverse axes of the NCS through Eqs. (\ref{TA}). The other is the intrinsic degree of freedom with respect to the NCS. They characterize the polarization in the following way.
On one hand, the SV fixes, via Eqs. (\ref{varpi}) and (\ref{QUT2}), a natural representation for the polarization. The wavefunction of polarization, the Jones function, in one particular natural representation is defined over the associated NCS. The Stokes parameters that the Jones function determines through the Pauli matrices (\ref{PM}) are specified with respect to the same NCS, forming the Stokes vector (\ref{SV}).
All the natural representations are equivalent in the sense that a polarization state can be described in any natural representation by a corresponding Jones function. The Jones functions in two different natural representations are transformed in accordance with Eq. (\ref{T-JV}) and the Stokes vectors are transformed in accordance with Eq. (\ref{T-SV}).
On the other hand, the intrinsic degree of freedom with respect to the NCS is represented by the Pauli matrices in the associated natural representation. What is surprising is that a given value of the intrinsic degree of freedom with respect to different NCS's does not mean the same polarization.
As a result, there are two independent mechanisms to change the polarization of a light beam.
One is to change the value of its intrinsic degree of freedom, or its Jones function, with its SV remaining fixed. The other is to change its SV, or its NCS, with its Jones function remaining fixed.
The former can be expressed mathematically as an arbitrary SU(2) rotation of the Jones function in a fixed NCS. What is performed by use of polarizers \cite{Dama} falls into that category.
The latter takes the form of a SO(3) rotation of the polarization vector about the momentum by an angle $\Phi$ that satisfies Eq. (\ref{R-LTA1}). This is the case with all the recent experiments \cite{Zhan} that convert uniformly polarized paraxial beams to radially or azimuthally polarized paraxial beams.

It is further inferred from the correspondence (\ref{HO}) that there are an infinite number of eigenstates of the helicity belonging to the same eigenvalue. They are described by the same eigenfunction of the Pauli matrix $\hat{\sigma}_3$ in different natural representations. Each of them shows a distinct state of polarization though they are only different by a helicity-dependent phase. It is this phase that accounts for the geometric phase in the paraxial approximation \cite{Host-K}.
By the way, what Hosten and Kwiat observed in their experiments is the effect of the phase of the polarization vector in Eq. (\ref{a3-P}), instead of the effect of the phase of the polarization wavefunction in Eq. (\ref{RT-alpha3}).

It should be emphasized that the new degree of freedom, the SV, which has observable physical effects, is different from what we are familiar with. It is not an observable by itself.
What it conveys is how to determine, relative to the GCS, the NCS with respect to which the physically observable quantity, the intrinsic degree of freedom, is specified.
Based on this conclusion, it is not difficult to imagine that only in one particular natural representation can the polarization of the radiation field be strictly quantized as is suggested by the canonical quantization condition (\ref{CCR}).
It is hoped that the findings presented here will deepen understanding of the quantum properties of the radiation field.

\end{document}